\documentclass{llncs}
\usepackage{graphicx}
\usepackage{enumerate}
\usepackage{amsmath}
\usepackage{amsfonts,amssymb}
\usepackage{algorithm}
\usepackage{algorithmic}

\begin{document}
\title{Personalized Classifier Ensemble Pruning Framework for Mobile Crowdsourcing}
\author{
Shaowei Wang\inst{1} \and
Liusheng Huang\inst{2} \and
Pengzhan Wang\inst{3} \and
\\Hongli Xu\inst{4} \and
Wei Yang\inst{5}
}
\institute{School of Computer Science and Technology, \\University of Science and Technology of China, Hefei 230027, Anhui, China \\
\email{\{wangsw\inst{1}, pzwang\inst{3}\}@mail.ustc.edu.cn,\\\{lshuang\inst{2}, xuhongli\inst{4}, qubit\inst{5}\}@ustc.edu.cn}
}

\maketitle

\begin{abstract}
Ensemble learning has been widely employed by mobile applications, ranging from environmental sensing to activity recognitions. One of the fundamental issue in ensemble learning is the trade-off between classification accuracy and computational costs, which is the goal of ensemble pruning. During crowdsourcing, the centralized aggregator releases ensemble learning models to a large number of mobile participants for task evaluation or as the crowdsourcing learning results, while different participants may seek for different levels of the accuracy-cost trade-off. However, most of existing ensemble pruning approaches consider only one identical level of such trade-off. In this study, we present an efficient ensemble pruning framework for personalized accuracy-cost trade-offs via multi-objective optimization. Specifically, for the commonly used linear-combination style of the trade-off, we provide an objective-mixture optimization to further reduce the number of ensemble candidates. Experimental results show that our framework is highly efficient for personalized ensemble pruning, and achieves much better pruning performance with objective-mixture optimization when compared to state-of-art approaches.
\end{abstract}
\keywords{Ensemble Pruning, Mobile Crowdsourcing, Bi-objective Optimization, Pareto Optimization}

\section{Introduction}\label{sec:intro}
A classifier ensemble leverages a bundle of diverse base classifiers (e.g., SVM, k-NN, Bagging classifier, Boosting classifier) to complete classification tasks by aggregating the classification results of the base classifiers. Since individual classifiers may perform poorly whereas the ensemble of these classifiers usually significantly improves, classifier ensembles are popular and are widely adopted in mobile device applications, ranging from environment-centric scenarios (e.g., place classification\cite{zhu2013feature}) to people-centric scenarios (e.g., activity recognition \cite{catal2015use}, human fall detection \cite{gibson2016multiple}\cite{sun2015iprotect}).

Though classifier ensemble could potentially improve classification accuracy, evaluating every member in the ensemble is much more expensive than the evaluation of an individual classifier. On the other hand, the number of classifiers in the ensemble limits the achievable classification accuracy. Such dilemma calls for the trade-off between classification accuracy and the classification cost (e.g., the size of the ensemble or time and memory consumptions of the ensemble) for classifier ensembles. Hence, \textit{ensemble pruning} \cite{zhou2002ensembling} has been proposed to select a subset of all possible classifiers meanwhile avoid harming classification accuracy. It's worth noting that the accuracy-cost trade-off is especially important in mobile crowdsourcing scenarios, where mobile devices' power saving is a crucial issue and realtime online classification is usually needed.

Specifically, in the crowdsourcing scenarios, the centralized crowdsourcing server may need to publish pruned classifier ensembles to every crowdsourcing participants, mainly due to two types of purposes. One purpose is facilitating each participant to finish crowdsourcing tasks, as many crowdsourcing tasks rely on ensemble classification. The other purpose is publishing the learned classifier ensembles as crowdsourcing result. This happens in reciprocal crowdsourcing tasks, where each participant contributes limited data (due to privacy consideration) to the centralized server, to collectively train a ensemble classification model, and then receive the learned classifier ensemble from the server, who has much more computation and storage resources than each single mobile devices for training the ensemble model. Consequently, each participant can then do classification tasks on their own mobile devices without sacrificing data privacy. In concise, during mobile crowdsourcing, the server may need to publish pruned classifier ensembles to plenty of participants that equipped with mobile devices.

We argue that participants need customized levels of accuracy-cost trade-off in ensemble pruning, as some participants may pursue better classification accuracy, while some other may pay more attention to energy saving or time saving during classification. One naive approach to tackle such personalized ensemble pruning challenge is running existing ensemble pruning approaches multiple times with different trade-off parameters, but this would be too costly especially when the number of participants is relatively large.

Motivated by resilience of bi-objective optimization, which simultaneously optimizes for two objectives, in this paper, we embed personalized ensemble pruning problem into bi-objective Pareto optimization problem by taking the classification accuracy and its cost as the objectives. We also propose an improvement in Pareto optimization by mixing up the accuracy and the cost objectives, such that the searching space of ensembles is concentrated when each participant's trade-off level is known by the server. More specifically, this work makes the following contributions:
\begin{itemize}
         \item[\textbf{--}] We formulate the personalized ensemble pruning problem, which captures the need of personalized accuracy-cost trade-off levels of crowdsourcing participants.
         \item[\textbf{--}] We present a personalized ensemble pruning framework based on bi-objective Pareto optimization, we further provide an objective-mixture improvement to Pareto optimization for linear-combination loss function.
         \item[\textbf{--}] We conduct extensive experiments to validate the efficiency and effectiveness of our framework.
\end{itemize}

The remainder of this paper is organized as follows. Section \ref{sec:model} provides preliminary knowledge about bi-objective optimization and ensemble pruning. Section \ref{sec:frame} gives our personalized ensemble pruning model and framework. Section \ref{sec:opts} describes the objective-mixture improvement on Pareto optimization. Section \ref{sec:anal} theoretically analyzes our framework. Section \ref{sec:exp} presents experimental results. Section \ref{sec:rel} reviews related work on ensemble pruning. Lastly section \ref{sec:conc} concludes the whole paper.

\section{Preliminaries}\label{sec:model}
In this section, we briefly introduce bi-objective Pareto optimization and the formal definition of ensemble pruning.

\subsection{Bi-objective Pareto Optimization}
Multi-objective optimization explicitly optimizes for two or more objectives and outputs multiple solutions, in which each solution has at least one advantage objective when compared to any other solutions. Conventionally the \textit{Pareto optimal} rule is used to determine the advantage or domination relation between solutions, and a solution is Pareto optimal if it is not dominated by any other solutions. Specifically, the Pareto domination relation in bi-objective optimization is formulated as follows:
\begin{definition}[Pareto Dominance]\label{def:domi}
Let $f=(f_1,f_2): \Omega \rightarrow \mathbb{R}^2$ be the objective vector. For two solution $S_1, S_2 \in \Omega$, $S_1$ dominates $S_2$ iff $f_1(S_1)\leq f_1(S_2)$ and $f_2(S_1)\leq f_2(S_2)$, and $f_1(S_1)<f_1(S_2)$ or $f_2(S_1)<f_2(S_2)$ holds.
\end{definition}

Unlike one-objective optimization that usually only have one optimal solution, bi-objective optimization returns a set of Pareto optimal solutions. This resilience property of bi-objective optimization lately motivates our framework to use bi-objective optimization to simultaneously approximate optimal classifier ensembles with diverse trade-off levels.

\subsection{Ensemble Pruning}

Ensemble pruning selects a subset of classifiers $S$ from a pool of base classifiers $T$, such that the accuracy of aggregated classification results only slightly degrades or even improves, meanwhile the number of classifiers (or computational costs) in $S$ is significantly reduced compared to the whole classifier pool $T$.

To be precise, let the number of base classifiers in $T=\{t_0,..., t_m\}$ be $m$, and selected classifiers $S\in\{0,1\}^m$ be a boolean vector, in which each bit indicates whether each $t_i$ is selected. Ensemble pruning seeks for a trade-off between the classification error rate $E(S)$ on a validation data set and the cost to evaluate classifiers in $S$ as follows:
\begin{equation}\label{eq:ep}
S_{opt} = \text{arg min}_{S\in\{0,1\}^m}\ E(S)+\alpha\cdot|S| ,
\end{equation}
Where $\alpha \in [0.0, +\infty)$ is the trade-off level, $|S|=\sum_{k=1..m}{s_i}$ is number of pruned classifiers, and $E(S)+\alpha\cdot|S|$ is the combined loss function. Without loss of generalization, we treat the evaluation cost of each classifier as $1$ unit here, but one can also use the actual time or(and) memory consumption of evaluating each classifier on mobile devices as the computational costs.

\section{Model and Framework}\label{sec:frame}

\subsection{Personalized Ensemble Pruning Model}
Recall that during crowdsourcing, the server usually need to publish pruned classifier ensembles to crowdsourcing participants. In practical, some crowdsourcing participants may pursue best achievable classification accuracy for better quality of services, while some others may pay more attention to energy saving or time saving during classification. This implies that ensemble pruning with personalized accuracy-cost trade-off levels is needed in mobile crowdsourcing scenarios.

\begin{figure}[htbpp]
\centering
\includegraphics[width=80mm]{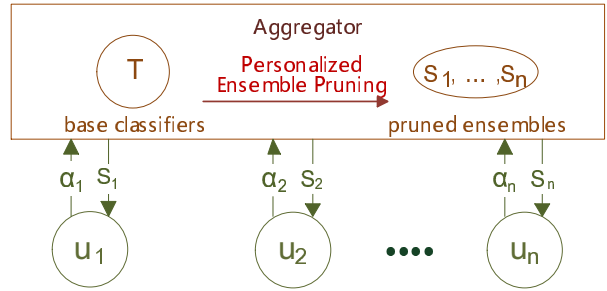}
\vspace*{-1ex}
\caption{Publishing personally pruned classifier ensembles in crowdsourcing.}
\vspace*{-1ex}
\label{fig:model}
\end{figure}

Formally, supposing that there are $n$ crowdsourcing participants, each has a choice of the accuracy-cost trade-off level $\alpha_j$ for ensemble pruning, then the optimal pruned ensemble $S_{j,opt}$ for participant $u_j$ can be formulated as follows:
\begin{equation}\label{eq:pep}
S_{j,opt} = \text{arg min}_{S\in\{0,1\}^m}\ E(S)+\alpha_j\cdot|S|.
\end{equation}
Consequently, the aggregator should publish personalized optimal classifier ensemble $S_{j,opt}$ to the participant $u_j$, as showed in Fig.\ref{fig:model}. But practically, as finding one optimal classifier ensemble $S_{j,opt}$ in exponential-sized search space is already computationally hard, a approximation $S_j$ of $S_{j,opt}$ is preferable instead.

The main challenge for personalized ensemble pruning is how to efficiently approximate all $S_{j,opt}$ for each participant when the number of participants is relatively large.

\subsection{Personalized Ensemble Pruning Framework}
We now provide our basic personalized ensemble pruning framework based on bi-objective optimization. Since finding the optimal pruned ensemble for each participant with trade-off level $\alpha_j$ would be too costly, especially when the number of participants $n$ is large, here we resort to bi-objective Pareto optimization, which takes the empirical classification error $E(S)$ on validation data set as objective $f_1(S)$ and computational costs $|S|$ as objective $f_2(S)$ in Definition \ref{def:domi}, and returns a Pareto optimal ensemble set $P=\{p_1, p_2, ..., p_l\}$ as solutions, where $l$ is size of the Pareto optimal solution set.

\begin{algorithm}
    \caption{Personalized Ensemble Pruning Framework}
    \label{alg:framework}
    \begin{algorithmic}[1]
        \REQUIRE A set of trained base classifier $T=\{t_1, t_2, ..., t_m\}$, \\ \ \ \ \ \ \ participants' trade-off levels $A=(\alpha_1, \alpha_2, ..., \alpha_n)$.
        \ENSURE Personalized pruned ensembles $\mathbb{S}=(S_1, S_2, ..., S_n)$ for each participant.
        \STATE Let $f(S)=(E(S), |S|)$ be the bi-objective.
        \STATE $P= \text{bi-objective-solver}(f(S))$
        \FOR{$j=1$ \TO $n$}
            \STATE $S_{j} = \text{arg min}_{S\in P}\ E(S)+\alpha_j\cdot|S|$
        \ENDFOR
        \RETURN $\mathbb{S}=(S_1, S_2, ..., S_n)$
    \end{algorithmic}
\end{algorithm}

As detailed in Algorithm \ref{alg:framework}, in our framework, after a tractable Pareto ensemble set $P$ is returned by bi-objective solver,  the approximate personalized optimal pruned ensemble $S_j$ is selected from $P$. As long as the Pareto optimal set $P$ is sufficiently optimized and complementary, the personalized pruned ensemble $S_j$ for participant $u_j$ will be a good approximation to $S_{j,opt}$ in Formula \ref{eq:pep}.

\begin{figure}[htbpp]
\centering
\includegraphics[width=90mm]{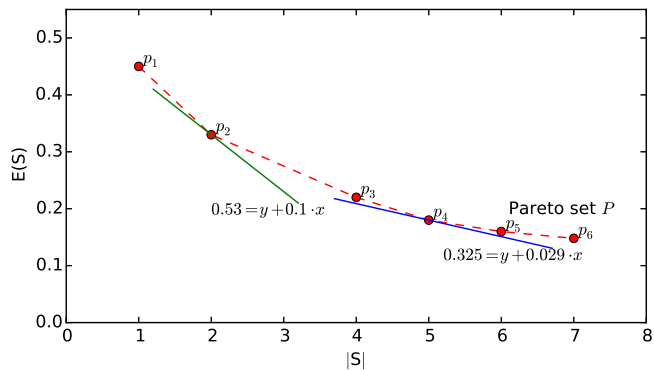}
\vspace*{-1ex}
\caption{An example of selecting personalized pruned ensembles from the Pareto set $P$.}
\label{fig:example}
\end{figure}

In Fig. \ref{fig:example}, we demonstrate how the framework selects personalized pruned ensembles for two participants with trade-off levels $0.1$ and $0.029$ from Pareto ensemble set $P$. In this case, the tractable minimum combined loss in Formula \ref{eq:pep} for participant with trade-off level $0.1$ is $0.534$, for participant with trade-off level $0.029$ is $0.325$.

Naively, selecting optimal ensemble for each participants from Pareto set in line $3-5$ or in the Fig. \ref{fig:example} has computational complexity of $n\cdot l$, which is inefficient when number of participants is relatively large, where $l$ is size of $P$ and is related to number of iterations in the bi-objective solver, hence $l=O(poly(m))$. Fortunately, by sorting the trade-off levels in $A$ and the Pareto set $P$ as showed in Algorithm \ref{alg:selection}, the complexity can be optimized.

\begin{algorithm}[H]
    \caption{Personalized Ensemble Selection}
    \label{alg:selection}
    \begin{algorithmic}[1]
        \REQUIRE The Pareto ensemble set $P=\{P_1, P_2, ..., P_l\}$, \\ \ \ \ \ \ \ participants' trade-off levels $A=(\alpha_1, \alpha_2, ..., \alpha_n)$.
        \ENSURE Optimal pruned ensembles $\mathbb{S}=\{S_1, S_2, ..., S_n\}$ in $P$ for each participant.
        \STATE sort $P$ by $|P_i|$ in descending order
        \STATE sort $A$ in ascending order
        \STATE $i=1$
        \FOR{$j=1$ \TO $n$}
            \WHILE{$i< l$ and $E(P_{i+1})+\alpha_j\cdot |P_{i+1}| \leq E(P_i)+\alpha_j\cdot |P_i|$}
                \STATE $i=i+1$
            \ENDWHILE
            \STATE $S_j=P_i$
        \ENDFOR
        \RETURN $\mathbb{S}=\{S_1, S_2, ..., S_n\}$
    \end{algorithmic}
\end{algorithm}

Note that any algorithm solves the bi-objective Pareto optimization problem can be deployed in our framework. Recently, Qian \textit{et al.}\cite{qian2015pareto} proposed an evolutionary algorithm for non-personalized ensemble pruning with guaranteed theoretical advantages over former approaches, this outstanding work on non-personalized ensemble pruning can naturally be embedded into our framework for personalized ensemble pruning.

\section{Objective-Mixture Optimization}\label{sec:opts}
Recall that in our basic framework, the bi-objective $f(S)=(E(S), |S|)$ solver returns a set of Pareto optimal ensembles $P=\{p_1, p_2, ..., p_l\}$, among which for any $p_i, p_j \in P$, $p_i$ doesn't dominate $p_j$. However, when taking the concrete value of participants' trade-off levels $A=\{\alpha_1, \alpha_2, ..., \alpha_j\}$ into consideration, the non-dominate relation between $p_i$ and $p_j$ may not hold for any level $\alpha \in A$, formally we have:
\begin{equation}\label{eq:basicdomi}
(E(p_i), |p_i|) \not\geq (E(p_j), |p_j|) \ \nRightarrow \ (E(p_i)+\alpha\cdot|p_i|) \not\geq (E(p_j)+\alpha\cdot|p_j|).
\end{equation}
This implies that the Pareto optimal ensemble set $P=\{p_1, p_2, ..., p_l\}$ still has redundances when concrete trade-off levels $A=\{\alpha_1, \alpha_2, ..., \alpha_j\}$ or their ranges are already known. Consequently, some redundant $p_i\in P$ will not be selected as pruned ensemble by any participants, but will mislead optimization in the bi-objective solving process. In order to mitigate such redundances, we tweak the bi-objective $f(S)$ in line $1$ of Algorithm \ref{alg:framework} as follows:
\begin{equation}\label{eq:optbi}
f(S)=(E(S)+\alpha_{min}\cdot |S|, E(S)+\alpha_{max}\cdot |S|),
\end{equation}
where $\alpha_{min} = min \{\alpha_1, \alpha_2, ..., \alpha_j\}$, $\alpha_{max} = max \{\alpha_1, \alpha_2, ..., \alpha_j\}$.

As a result, with this optimized objective-mixture framework, for any $p_i, p_j \in P$, there exists an $\alpha\in [\alpha_{min},\alpha_{max}]$ that the following formula holds:
\begin{equation}\label{eq:optdomi}
\begin{aligned}
&(E(p_i)+\alpha_{min}\cdot |p_i|, E(p_i)+\alpha_{max}\cdot |p_i|) \not\geq (E(p_j)+\alpha_{min}\cdot |p_j|, E(p_j)+\alpha_{max}\cdot |p_j|)\\ &\Rightarrow (E(p_i)+\alpha\cdot|p_i|) \not\geq (E(p_j)+\alpha\cdot|p_j|).
\end{aligned}
\end{equation}

Roughly speaking, any $p_i\in P$ in this optimized framework with objective-mixture has the potential to be finally selected as pruned ensemble by a participant.

Practically, when $\alpha_{min}$ and $\alpha_{max}$ is too close, which may cause bi-objective optimization converges local optimally, some additive or multiplicative relaxations could be putted on $\alpha_{min}$ and $\alpha_{max}$, which means the relaxed objectives is $f(S)=(E(S)+c_{min}\cdot\alpha_{min}\cdot |S|, E(S)+c_{max}\cdot\alpha_{max}\cdot |S|)$, where $c_{min}\leq 1.0, c_{max}\geq 1.0$ are the multiplicative relaxation factors.

\section{Performance Analyses}\label{sec:anal}
In this section, we analyse the theoretical computational complexity and ensemble pruning performance of our personalized ensemble pruning framework.

Computational complexity of our framework highly depends on the deployed bi-objective solver at line $2$ in Algorithm \ref{alg:framework}. Fortunately, the evolutionary bi-objective solver in \cite{qian2015pareto} shows that expected $O(m^2\log m)$ iterations is sufficient for approximating optimal Pareto set, where $m$ is the number of base classifiers.

The personalized pruned ensemble selection in line $3-5$ naively has computational complexity of $n\cdot l$, but in Algorithm \ref{alg:selection}, its complexity is reduced to $O(n\log n + l\log l)$, where $l$ is size of $P$ and $l=O(poly(m))$.

To put it concisely, the overall expected computation complexity of our framework is $O(poly(m) + n\log n)$, while the naive ensemble pruning approach  for personalized ensemble pruning tends to cost $\Theta(poly(m)\cdot n)$. Hence our framework is efficient in terms of the number of participants $n$.

The gap between selected pruned ensemble for each participant in our framework and the optimal ensemble in Formula \ref{eq:pep} also highly depends on the deployed algorithm of bi-objective solver. The evolutionary bi-objective solver in \cite{qian2015pareto} has been proved to outperform former ensemble pruning approaches, thus the achievable performance of our framework is guaranteed by \cite{qian2015pareto}.

\section{Experiments} \label{sec:exp}
In this section, we evaluate the computational overheads and pruning performance of our personalized ensemble pruning framework, with comparison to the state-of-art ensemble pruning approach PEP in \cite{qian2015pareto}, which is also used by our framework as the bi-objective solver at line $2$ in Algorithm \ref{alg:framework} (the differences between PEP and our framework is declared in the next section).

We mainly compare the running time and combined loss function (in Formula \ref{eq:pep}) performance of personalized ensemble pruning in our basic framework (BF) and the objective-mixture framework (OMF) with the approach PEPs that runs the PEP multiple times to get personalized pruned ensembles.

In order to simulate ensemble pruning for crowdsourcing participants that equipped with mobile devices, the human activity recognition data set based on mobile phone accelerometers readings from \cite{kwapisz2011activity} is used for our experiments. The data set has $4944$ examples with $43$ attributes extracted from time-serial data of mobile phone accelerometers, includes six classes of human activities: walking, jogging, upstairs, downstairs, sitting and standing.

In our experiments, the dataset is randomly split into tree parts: $60\%$ as training set, $20\%$ as validation set and $20\%$ as test set. The pool of $m=20$ base CART tree \cite{breiman1984classification} classifiers are generated by Bagging \cite{breiman1996bagging} on the training set. In these experiments, we simulated with number of participants $n$ ranges from $1$ to $1280$, with randomized personalized trade-off level $\alpha_j\in A$ ranges from $0.01$ to $0.2$, and the empirical multiplicative objective relaxation factors in OMF are empirical values $c_{min}=0.3, c_{max}=1.7$. The number of iterations of the bi-objective solver is set as $80$. We conduct experiments on the scikit-learn \cite{pedregosa2011scikit} platform.

\begin{figure}[H]
\centering
\includegraphics[width=110mm]{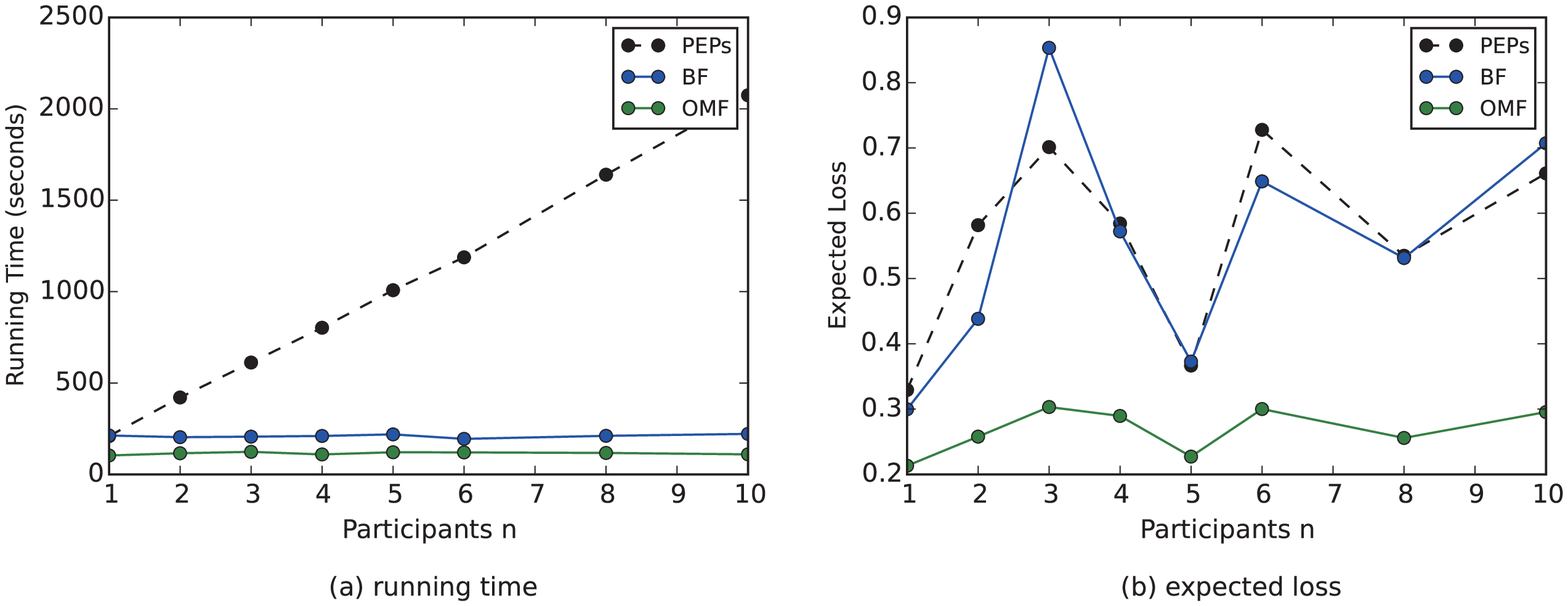}
\vspace*{-4ex}
\caption{The running time and expected loss (the smaller the better) of personalized ensemble pruning for $\{1, ..., 10\}$ participants.}
\label{fig:timeloss}
\end{figure}

The running time results with a relatively small number of participants are showed in Fig. \ref{fig:timeloss} (a). The PEPs approach for personalized ensemble pruning costs linear to $n$ computational time, while the time consumed by our BF and OMF approach only slightly increase with the number of participants $n$ increases. Besides, our framework with objective-mixture (OMF) averagely reduced $40\%$ running time than the basic framework (BF), mainly due to filtering out unnecessary ensemble candidates during each iteration. In concise, our framework for personalized ensemble pruning is efficient when the number of participants $n$ increases.

\begin{figure}[!t]
\centering
\includegraphics[width=100mm]{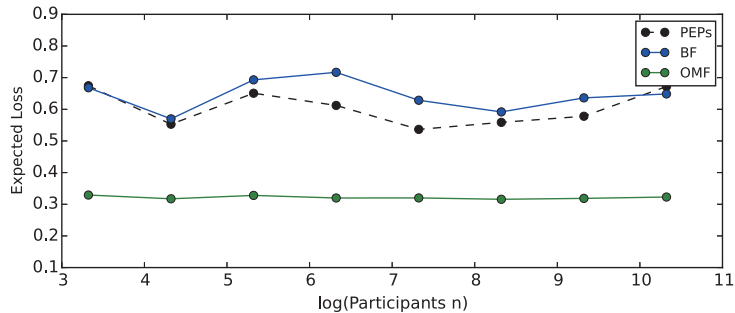}
\vspace*{-3ex}
\caption{The expected loss (the smaller the better) of personalized ensemble pruning for $\{10,20,40, ..., 640, 1280\}$ participants.}
\vspace*{-2ex}
\label{fig:loss}
\end{figure}

The expected loss function (in Formula\ref{eq:pep}) results of pruned ensembles for small number of participants are showed in Fig. \ref{fig:timeloss} (b), it demonstrates that there is no significant difference of loss function between the BF and the PEPs approach, while our framework with objective-mixture (OMF) reduced losses by $50\%$ when compared to the BF and PEPs approach. Combining experimental results in Fig. \ref{fig:timeloss} (a) as conclusion, our OMF approach outperforms the PEPs approach and our BF approach in both efficiency and effectiveness.

We also evaluate the performance of our approaches with relatively large number of participants, and the results showed in Fig. \ref{fig:loss}. The results demonstrate that the losses of pruned ensemble from objective-mixture OMF approach averagely outperforms the basic framework and the PEPs approaches by $50\%$.

\section{Related Work}\label{sec:rel}
Ensemble learning is widely used for data classification, in the form of Bagging\cite{breiman1996bagging}, Boosting\cite{freund1997decision} or heterogeneous classifiers (e.g. combination of SVM, k-NN and Naive Bayes classifiers). Mainly to mitigate the inefficiency issue in ensemble learning, ensemble pruning has been extensively studied.  As ensemble pruning itself is computational hard, genetic algorithms\cite{zhou2002ensembling} and semi-definite programming\cite{zhang2006ensemble} has been proposed.

Demir \textit{et al.}\cite{demir2005cost} and Ula{\c{s}} \textit{et al.}\cite{ulacs2009incremental} also consider costs of computations of ensemble or discriminants between classifiers to heuristically compose a pruned ensemble. Some other ensemble pruning approaches such as Margineantu  \textit{et al.} \cite{margineantu1997pruning} and Mart{\'\i}nez-Mu{\~n}oz  \textit{et al.} \cite{martinez2006pruning} are based on ordered aggregation, which rearrange the order of classifiers in base classifier pool. These approaches pruning ensemble with only one identical trade-off level and are hard modifying for personalized ensemble pruning.

Recently, Qian \textit{et al.} \cite{qian2015pareto} proposed an ensemble pruning approach using evolutionary bi-objective optimization, and proves its theoretical advantages over former heuristic or ordered based approaches. However, it doesn't considers personalized ensemble pruning for crowds nor the objective-mixture optimization when the personalized trade-off levels are known by the centralized aggregator. The concurrent work \cite{fan2016poster} considers customized accuracy-cost trade-off on a group of Pareto optimal sets, whereas this work focus improving efficiency and effectiveness of personalized trade-off on one Pareto optimal set.

Another topic related to multiple ensembles selection is \textit{dynamic ensemble selection} \cite{ko2008dynamic}, which dynamically select a classifier ensemble for one class of input samples, instead of for personalized accuracy-cost trade-off levels.

\section{Conclusion} \label{sec:conc}
In this paper, we formulate the problem of personalized classifier ensemble pruning in mobile crowdsourcing, and proposed a framework for the problem based on bi-objective Pareto optimization with objective mixture. In our framework, each participant (the ensemble user) could choose a customized level of trade-off between the ensemble's classification accuracy and its computational costs, so that the need of energy saving on mobile devices and the need of classification accuracy are balanced for each participant. Both theoretical and experimental results demonstrate the efficiency and effectiveness of our framework. Specifically, the experimental results show that our personalized ensemble pruning framework with objective mixture averagely reduced $40\%$ running time and $50\%$ combined losses simultaneously under same number of iterations.


\bibliographystyle{splncs03}
\bibliography{refs}
\end{document}